\documentstyle[11pt]{article}
\def\begref{\begingroup\let\INS=N}
\def\bref{\goodbreak\if N\INS\let\INS=Y\vbox{
\centerline{References}}\fi\hangindent\parindent
\hangafter=1\noindent\ignorespaces}
\def\begcap{\begingroup\let\INS=N}
\def\cap{\goodbreak\if N\INS\let\INS=Y\vbox{
{\bf Figure Captions}}\fi\hangindent\parindent
\hangafter=1\noindent\ignorespaces}
\def\s{\sigma}
\def\o{\omega}
\def\O{\Omega}
\def\a{\alpha}

\def\d{\delta}

\def\l{\lambda}
\def\G{\Gamma}
\def\g{\gamma}

\def\beq{\begin{equation}} 
\def\eeq{\end{equation}} 

\newcount\ec \ec=1 
\def\no{\eqno(\the\ec)\global\advance\ec by 1}
\def\eno{(\the\ec)\global\advance\ec by 1}

\newcount\nc \nc=1 
\def\in{\the\nc\global\advance\nc by 1}

\def\frac#1#2{{#1\over #2}}

\newcount\sectcount \sectcount=1
\def\sect#1{\bigbreak{{\noindent \s \the\sectcount \ \ #1}}
\global\advance\sectcount by 1\medskip\nobreak}

\def\pst#1{\bigbreak{{\noindent{\bf \the\sectcount. #1}}}
\global\advance\sectcount by 1\medskip\nobreak}
\begin{document}

\noindent {\Large\bf Stimulated  Raman scattering of water maser lines in
astrophysical plasmas}
\vskip 1 cm 
\noindent
{\bf R. T. Gangadhara}

\noindent
Indian Institute of Astrophysics, Bangalore--560034, India
\vskip 0.3 cm 
\noindent
{\bf Shuji Deguchi}

\noindent
Nobeyama Radio Observatory, National Astronomical Observatory,
 Minamimaki, Minamisaku, Nagano 384-13, Japan
\vskip 0.3 cm 
\noindent
{\bf H. Lesch}

\noindent
Institut f\"ur Astronomie und Astrophysik der Universit\"at M\"unchen,
Scheinerstra$\beta$e 1, 81679 M\"unchen, Germany

\vskip 1 cm 

Radiative transfer equations are derived and solved for the stimulated
Raman scattering of water maser lines in the astrophysical plasmas with
electron density of about $10^6$-$10^7$ ${\rm cm^{-3}}.$ In stimulated
Raman scattering, the energy of water maser line is transferred to the
side band modes: Stokes mode and anti-Stokes mode. The Stokes mode is easily
produced by backward Raman scattering while the anti-Stokes mode is created by
the interacting intersecting masers in the plasma. The intensity of the Stokes mode
is higher than that of the anti-Stokes mode. These side band modes
are proposed as explanation for the extreme velocity features observed in the
galaxy NGC~4258. The threshold value of the brightness temperature for the
Raman scattering is about $10^{16}-10^{19}$~K, and it is satisfied in the
case of NGC~4258. 
\vskip .5 cm 



\vfill\eject
 \noindent{\large \bf I.\ \  INTRODUCTION}\par\noindent
        Stimulated scattering processes become important in astrophysical
plasmas when the incident radiation on plasma has a very high brightness 
temperature $(T_b>10^{16}$~K) $^{1,2,3,4}.$
In practice, line radiation of water masers in the galactic nuclei 
and continuum radiation from quasars are considered to have such an high 
brightness. Stimulated Raman and Compton scatterings will occur in the high 
and low density plasmas, respectively, depending upon whether the collective 
effect of electrons works or not $^5.$ Though a 
number of studies on the stimulated Raman scattering have been made, the transfer 
of line radiation with shifted frequency is not well understood. 
Fernandez and Reinnisch$^6$ have made an attempt made to formulate the theory of
stimulated Raman scattering with radiative transfer equation for specific
intensities.
\par
     In this paper, we have made an attempt study stimulated Raman scattering 
of water maser line with shifted frequency. A practical application of this
theory is aimed to explain the extreme high-velocity features of water maser 
lines at 22.235 GHz in the galactic nucleus of NGC~4258. The high-velocity 
features of this line appears separated  by 1000~km~s$^{-1}$ from the main 
component of the water maser line. Though the recent Very Long Baseline
Interferometry (VLBI) observations showed 
some evidence of this component arising from the rotating disk, all the 
observations (for example, the weakness of the  components) are not 
necessarily well explained by the rotating disk model $^7.$ The beaming of
Raman maser is discussed in Sec.~II. The radiative transfer equations for stimulated
Raman scattering are derived and solved in Sec.~III.
Further we discuss the limitation and applicability of our model to the 
astrophysical maser phenomenon in Sec.~IV.
\vskip 0.3 truecm
\noindent{\large \bf II.\ \ BEAMING OF RAMAN MASERS } \par\noindent
The principal mechanism behind the beaming of astrophysical masers is the 
frequency change of the resonance radiation. As usual for astrophysical 
masers, the Raman masers are also expected to be strongly beamed. 
The masers are amplified in the direction with smallest gradient in the
resonance frequency where the optical depth of the resonance radiation
becomes maximum. For the case of Raman masers, the change in the resonance
frequency is produced by two effects: change of plasma frequency due
to temperature and density variations, and Doppler effect due to the
gas flow in plasma.
\par
        The frequencies of incident and scattered waves are related by
$$\o_1=\o_o-\o_3,\eqno (1)$$
where $\o_o$, $\o_1$ and $\o_3$ are the frequencies of incident, scattered
and longitudinal waves, respectively. The frequency of the longitudinal wave is
determined by the dispersion relation
        $$      \o_3^2=\o_{pe}^2+3v^2_tk_3^2,      \eqno (2)$$
where $v_t^2=k_BT/m_e$ and $\o_{pe}^2=4\pi N_e e^2/m_e,$ $e$ and $m_e$ are the charge
and mass of electron, $N_e$ and $T$ are the density and temperature of electron plasma,
and $k_B$ is the Boltzmann constant.
\par
        For the case $\o_3\ll\o_o$, we can approximate
$k_3=2\o_o\cos\theta_3/c$ and $\theta_3=(\pi-\theta_1)/2,$  where 
$\theta_3$ is the angle between $\vec k_o$ and $\vec k_3$, and 
$\theta_1$ is the angle between $\vec k_o$ and $\vec k_1,$ and $c$ is the speed of
light. Therefore, the dispersion relation can be transformed as
        $$\o_3^2=\o_{pe}^2+6 (v_t^2/c^2)\o_o^2(1-\cos\theta_1).\eqno (3)$$
\par
        In astrophysical plasmas, density and temperature of the plasma
are not constant in space, and hence 
the resonance frequency varies along the propagation path of the waves.
\par
        In addition, the Doppler effect due to the gas flow in plasma
changes the frequency, which can be expressed in the rest frame as
$$\o_1=[\o_o(1+\vec v .\hat{n}_o/c)-\o_3](1-\vec v.\hat{n}_1/c),\eqno(4)$$ 
where $\vec v$ is the velocity of the plasma in
the rest frame, and $\hat{n}_o$ and $\hat{n}_1$ are the unit vectors
representing the propagation directions of incident and scattered waves,
respectively.
\par
        The frequency shift along the co-ordinate $s,$ which is taken in the 
direction of scattered wave, can be expressed as
$$\frac{d\o_1}{ds}=\o_o(\hat{n}_1.\nabla)(\vec v.\hat{n}_o/c)-\o_{10}
  (\hat{n}_1.\nabla )(\vec v.\hat{n}_1/c)- (\hat{n}_1.\nabla)\o_3,\eqno (5)$$
where $\o_{10}$ is the original value of frequency $\o_1.$ The first two
terms on the right-hand-side represent the frequency change due to Doppler 
motion of the gas, and third term is due to the density and temperature 
changes in plasma.
\par
        The transfer of line radiation in the moving medium has been well
studied and an approximation of large velocity gradient
has been developed. For example, for a
spherically expanding cloud, the Doppler term (in this case $\o_o=\o_1)$
can be written as
$$\left (\frac{d\o}{ds}\right )_{\rm Doppler}=\frac{\o_o}{c}\frac{dv}{dr}
   (\cos\theta_1-1),\eqno (6)$$
where we have taken the $z$ axis parallel to the direction of incident
radiation, which is considered to be the direction of maximum optical depth
$(\theta_1=0).$ The frequency of the scattered
radiation is not influenced by the Doppler motion of the matter in the
case of pure-forward scattering.
\par
        The density-temperature gradient term can be written in unit of 
$(\o_o/c)$ $(dv/dr)$ as
$$
\left(\frac{d\o}{ds}\right)_{D-T}  = 
\frac{\o_o}{c}\frac{dv}{dr}[\cos\theta_1\{
a_z+b_z(1-\cos\theta_1)\}+\sin\theta_1\cos\phi_1\{a_x+b_x(1-\cos\theta_1)\} $$
$$ +\sin\theta_1\sin\phi_1\{a_y+b_y(1-\cos\theta_1)\}],\eqno (7) $$
where vectors $a$ and $b$ are the non-dimensional parameters, which are 
proportional to the density (a) and temperature (b) gradients.
\vskip 0.3 truecm
\noindent{\large \bf III.\ \  STIMULATED RAMAN SCATTERING OF MASER}\par\noindent
        The fundamental set of equations for the study of stimulated Raman
scattering in the plasma medium are
$$\frac{\partial N_e}{\partial t}+\nabla . (N_e\vec v)=0,\eqno (8)$$
$$\frac{\partial \vec v}{\partial t}+\vec v .\nabla \vec v=
-\frac{e}{m_e}\left (\vec E+\frac{1}{c}\vec v\times \vec B\right )-\frac{3}{N_e}v_T^2
\nabla N_e-\nu_c \vec v,\eqno (9)$$
$$\nabla .\vec E=4\pi e(N_i-N_e),\eqno (10)$$
$$\nabla\times\vec E=-\frac{1}{c}
        \frac{\partial \vec B}{\partial t},\eqno (11)$$
$$\nabla\times\vec B=\frac{4\pi}{c} \vec j+\frac{1}{c}
        \frac{\partial \vec E}{\partial t},\eqno (12)$$
The Eqs. (8) and (9) are the electron continuity and momentum equations.
respectively. The Eq.~(10) is the Poisson equation. The last two Eqs.
(11) and (12) are Maxwell's equations. Here $\vec E$ and $\vec B$ are the
electric and magnetic fields of the waves involved in the three-wave process.
\par
        Taking curl on the both sides of Eq.~(11) and substituting 
Eq.~(12) for $\nabla\times\vec B$, we get
$$\frac{\partial^2 \vec E}{\partial t^2}+c^2\nabla\times(\nabla\times\vec E)
  + 4\pi \frac{\partial \vec j}{\partial t}=0.  \eqno (13)$$
Let $N_e=N_o+\d N$ and $N_i=N_o$ be the electron and ion densities, 
respectively, where $\d N$ is the electron density perturbation induced by 
the ponderomotive force of the radiation field, and $N_o$ is the stationary
background neutralizing ion density.
\par
        The current density is defined as 
$$\vec j = e (N_i \vec v_i-N_e \vec v_e). \eqno (14)$$
Taking the time derivative of $\vec j$ and substituting into Eq.~(13), 
we get
$$\frac{\partial^2 \vec E}{\partial t^2}+c^2\nabla\times(\nabla\times\vec E)
  - 4\pi e N_o\frac{\partial \vec v}{\partial t}
  - 4\pi e \frac{\partial (\d N\vec v)}{\partial t}
  =0.  \eqno (15)$$
From Eq.~(10), we find
$$\d N = -\frac{1}{4 \pi e}\nabla . \vec E.  \eqno (16)$$
Using Eqs.~(9) and (16),  Eq.~(15) can be written as
$$
\frac{\partial^2 \vec E}{\partial 
t^2}+c^2\nabla\times(\nabla\times\vec E)
+\o_{pe}^2\vec E- 3 \frac{N_o}{N_e}v^2_T\nabla(\nabla . \vec E) = 
  - 4\pi e N_o (\vec v . \nabla\vec v+\nu_c\vec v) $$
$$-\frac{\o_{pe}^2}{c}
  (\vec v\times\vec B)+4\pi e \frac{\partial (\d N\vec v)}{\partial t}.\eqno (17) $$
Now, using $\vec v(\nabla .\vec v)=(1/2) \nabla v^2-\vec 
v\times(\nabla\times\vec v)$ and
$\nabla\times\vec v=e\vec B/(m_e c),$ $^8$, Eq.~(17)
can be written as 
$$\frac{\partial^2 \vec E}{\partial t^2}-c^2\nabla^2\vec E
-\frac{3}{2}\frac{N_o}{N_e}v^2_T\nabla(\nabla . \vec E)+\o_{pe}^2\vec E=
 - \frac{\partial (\vec v\nabla .\vec E)}{\partial t}-
2\pi e N_o \nabla v^2 - 4 \pi e N_o\nu_c \vec v.\eqno (18)$$
We use plane--wave approximation $^9$ such as 
$$\vec E (\vec r, t)=\frac{1}{2}\sum_{i=0-3}
\{\vec E_i(\vec r, t)e^{i(\o_i t-\vec k_i .\vec r)}+
\vec E_i^*(\vec r, t)e^{-i(\o_i t-\vec k_i .\vec r)}\}\eqno (19)$$
and a similar approximation for the electron motions:
$$\vec v(\vec r, t)=\frac{1}{2}\sum_{i=0-3}
\{\vec v_i(\vec r, t)e^{i(\o_i t-\vec k_i .\vec r)}+
\vec v_i^*(\vec r, t)e^{-i(\o_i t-\vec k_i .\vec r)}\}.\eqno (20)$$
In Eqs.~(19) and (20) the amplitudes $\vec E_i$ and
$\vec v_i$ are the slowly varying functions of space and time, where 
$i=0,$ 1, 2 and 3 stand for incident (pump) wave 
$(\vec k_o,\o_o),$ Stokes mode $(\vec k_1,\o_1),$ anti-Stokes mode $(\vec k_2, 
\o_2)$ and longitudinal plasma wave $(\vec k_3,\o_3),$ respectively.
\par
        The stimulated Raman scattering instability excites resonantly
when the following phase matching conditions are satisfied:
$$ \o_1  =  \o_o-\o_3,\quad\quad\vec k_1=\vec k_o-\vec k_3,\eqno (21a)$$
$$ \o_2 =\o_o+\o_3,\quad\quad\vec k_2=\vec k_o+\vec k_3. \eqno (21b) $$
\par
        The growth rate of instability is given by $^{10, 11}$
$$\g_o=\frac{k_3 v_o}{2}\left(\frac{\o_{pe}}{\o_1}\right)^{1/2},\eqno (22)$$
where $v_o=eE_o/(m_e\o_o)$ is the quiver velocity of the electrons in the 
incident radiation field.
\par
        Using the first-order approximation $^9$ to the electron 
motion we find the transfer equations for electric field amplitudes:
for the incident radiation (pump)
$$\frac{\partial E_o}{\partial t}+c \frac{\partial E_o}{\partial x_o}=
   -c_o E_o-\a (E_1 E_3\cos\psi_1-E_2E^*_3\cos\psi_2),\eqno (23)$$
for the Stokes mode
$$\frac{\partial E_1}{\partial t}+c \frac{\partial E_1}{\partial x_1}=
   -c_1 E_1+\a E_o E_3^*\cos\psi_1,\eqno (24)$$
for the anti-Stokes mode
$$\frac{\partial E_2}{\partial t}+c \frac{\partial E_2}{\partial x_2}=
   -c_2 E_2-\a E_o E_3^*\cos\psi_2\eqno (25)$$
and for the longitudinal wave
$$\frac{\partial E_3}{\partial t}+v_g \frac{\partial E_3}{\partial x_3}=
   -c_3 E_3+\a_3 (E_o E_1^*\cos\psi_1+E_2E_o^*\cos\psi_2),\eqno (26)$$
where $c_i=(\o_{pe}/\o_i)^2\nu_c\,\, (i=0,\,1,\,2,\,3)$ represents the free-free 
damping process, and the electron-ion collision frequency 
$\nu_c=4 \pi n_e e^4 \ln\Lambda/m_e^2 v_t^3.$ The coefficients are 
$\a=k_3e/(4 m_e\o_o),$ $\a_3=\o^2_{pe}\a/(\o_o\o_3).$
Here $v_g=3 v^2_Tk_3/\o_3$ is the group velocity of longitudinal wave and
$\psi_i$ ($i=1$ or 2) is the angle between $\vec E_o$ and $\vec E_1$ or 
$\vec E_2.$ The co-ordinates $x_i\,\, (i=0,\,1,\,2,\,3)$ represent the 
direction of propagation of each wave. 
\par
Only the frequency downshifted component (Stokes mode) is excited, and the upshifted
component (anti-Stokes) is not amplified at the general scattering angles even if it is
resonant $^{10,12}.$ It should be noted that the second terms on the right-hand-side of
Eqs.~(24) and (25) are plus and minus. This means that only the downshifted waves grow, but
the upshifted waves never grow. However the resonance can occur for either of the upshifted or 
downshifted components in general, but both the components appear only in the case of 
pure forward scattering.
\par
        For the longitudinal wave, Landau damping is more efficient than the 
        binary collisional damping. The Landau damping rate of the plasma wave
        is given by
$$\G=\left(\frac{\pi}{8}\right)^{1/2}\frac{(\o_{pe}\o_3)^2}
{(\mid\!\! k_3\!\!\mid v_T)^3}\exp[-\o_3^2/(2\mid\!\! k_3\!\!\mid v_T)^2].\eqno (27)$$
In the steady state, Eqs.~(23), (24) and (26) reduces to
$$\frac{\partial E_o}{\partial x_o}=
   -\frac{c_o}{c} E_o-\frac{\a}{c} E_1 E_3,\eqno (28)$$
$$\frac{\partial E_1}{\partial x_1}=
   -\frac{c_1}{c} E_o+\frac{\a}{c} E_o E_3^*,\eqno (29)$$
$$E_3=\frac{\a_3}{\G+c_3}E_oE_1^*, \eqno (30)$$
where we have assumed incident wave and scattered wave are polarized in the 
same direction $(\psi_1=0).$ Eliminating $E_3$ between the Eqs.~(28) and 
(30), we find
$$\frac{\partial E_o}{\partial x_o}=-\frac{c_o}{c} E_o-
\frac{\a\a_3}{c(\G+c_3)}\mid\!\! E_1\!\!\mid^2 E_o. \eqno (31)$$
The complex conjugate of Eq.~(31) is given by
$$\frac{\partial E_o^*}{\partial x_o}=-\frac{c_o}{c} E_o^*-
\frac{\a\a_3}{c(\G+c_3)}\mid\!\! E_1\!\!\mid^2 E_o^*. \eqno (32)$$
Multiplying Eq.~(31) by $E^*_o$ and Eq.~(32) by $E_o$, and adding, we get
$$\frac{\partial I_o}{\partial x_o}=-\frac{2c_o}{c} I_o-
    \frac{2\a\a_3}{c(\G+c_3)} I_1 I_o. \eqno (33a)$$
Similarly from Eq.~(29), we find
$$\frac{\partial I_1}{\partial x_1}=-\frac{2c_1}{c} I_1+
    \frac{2\a\a_3}{c(\G+c_3)} I_o I_1, \eqno (33b)$$
where $I_o=\mid\!\! E_o\!\!\mid^2$ and $I_1=\mid\!\! E_1\!\!\mid^2$ are the 
intensities of the incident
and scattered waves. In  Eq.~(33) the first term on the right hand
side represent the free-free collisional damping. While the second term in 
 Eq.~(33a) represents stimulated absorption of the incident radiation
and in   Eq.~(33b) it represents the stimulated emission of the
scattered radiation. 
\vskip 0.3 truecm
\noindent{\large \bf A.\ \  Solution of equation (33)}\par\noindent
        When the amplitudes of $I_o$ and $I_1$ are not too large, we solve 
 Eq.~(33) numerically. But when $I_o$ and $I_1$ are too large
we can ignore the collisional damping terms and find the analytical solutions. 
\par
     The direction of propagation of the incident and the scattered waves 
in the thick and the thin plasma media are shown in Fig.~1.
Let $\theta$ be the angle between the coordinates $x_o$ and $x_1$, then 
we have $x_o=x_1\cos\theta$, and  Eq.~(33) becomes
$$\frac{d I_o}{d s}=-\cos(\theta)\left[I_o+\frac{B_o}{A_o}I_1I_o\right], 
        \eqno(34a)$$
$$\frac{d I_1}{d s}=-\frac{A_1}{A_o}I_1+\frac{B_1}{A_o}I_1I_o, \eqno(34b)$$
where $d s=A_o dx_1,$ $A_o=2c_o/c,$ $B_o=\a\a_3/\{c(\G+c_3)\},$
$A_1=2c_1/c,$ and $B_1=B_o.$
\par
Defining $W_o=\log(I_o/I_c)$ and $W_1=\log(I_1/I_c)$, where $I_c$ is a 
normalization
constant, we write  Eq.~(34) as
$$\frac{d W_o}{d s}=-\cos(\theta)\left[1+\frac{B_o}{A_o}I_c 
\exp(W_1)\right], 
\eqno(35a)$$
$$\frac{d W_1}{d s}=-\frac{A_1}{A_o}+\frac{B_1}{A_o}I_c\exp(W_o). 
                \eqno(35b)$$
We numerically solve  Eq.~(35) using the plasma density 
$N_o=3\times 10^6$~cm$^{-3}$,
the electron temperature $T=4200$~K, frequency of the incident radiation 
$\nu_o=22.235$~GHz and $I_c=A_o/B_o$. 
\par
The incident and the scattered wave intensities as functions of $s,$ at 
different values of the scattering angle $\theta=120^o,$ $140^o,$ $160^o$ 
and $180^o$ in a thick plasma medium, are plotted in Fig.~2.
At $s=0,$ we started with initial conditions  $I_o/I_c=1$  and   
$I_1/I_c=10^{-30}.$   As   $s$ increases scattered wave 
intensity grows much faster than incident wave intensity, indicating transfer
of energy from the incident wave to the scattered wave. At large $s,$ say 3.6,
and $\theta=180^o,$ intensity of the incident wave becomes vary large 
consequently the scattered mode also becomes  much stronger. The range
of $s,$ over which the incident and the scattered waves interact, increases 
with decreasing $\theta.$
Similarly, Fig.~3 shows the incident wave and the scattered wave intensities
as functions of $s,$ at different values of the scattering angle $\theta=
120^o,$ $140^o,$ $160^o$ and $180^o$ in the case of thin plasma medium.
Again at $s=0,$ we started with initial conditions  $I_o/I_c=10^3$  and 
$I_1/I_c=10^{-30}.$   At large $s,$ say 1, and $\theta=180^o$ 
intensity of the incident wave becomes vary large consequently scattered mode also 
becomes much stronger. 
In Figs. 2 and 3, the incident wave ($I_0$) comes
from the right and escaping from the boundary at
$s=0,$ and scattered wave ($I_1$) comes from the
left at $s=0.$ The boundary of plasma at the
right side exists at an arbitrary point, for example, at $s=3$--$6$ in Fig.~2, or
$s \approx 1$ in Fig.~3.
\par
When the stimulated absorption of $I_o$ and emission of $I_1$ are very large
compared to free-free absorption rates we can ignore the collisional damping 
terms in  Eq.~(33), therefore, we have
$$\frac{d I_o}{d s}=-\cos(\theta)\frac{B_o}{A_o}I_1 I_o, \eqno(36a)$$
$$\frac{d I_1}{d s}=\frac{B_1}{A_o}I_o I_1. \eqno(36b)$$
Solutions of  Eq.~(36) can easily be obtained using the 
following convenient transformations$^{13}:$
$$U_o=\frac{B_1}{A_o}I_o,\quad\quad  
U_1=\cos(\theta)\frac{B_o}{A_o}I_1.\eqno(37)$$ 
Now,  Eq.~(36) becomes
$$\frac{d U_o}{d s}=-U_1 U_o, \eqno(38a)$$
$$\frac{d U_1}{d s}=U_o U_1. \eqno(38b)$$
In the case of forward Raman scattering, we can define a constant of motion
$$U_o(s)+U_1( s)=m=U_o(0)+U_1(0),\eqno(39)$$
the solutions of  Eq.~(38) can then be written as
$$U_o(s)=\frac{m U_o(0)}{U_o(0)+U_1(0)\exp(m s)},\eqno (40a)$$
$$U_1(s)=\frac{m U_1(0)}{U_1(0)+U_o(0)\exp(-m s)}.\eqno (40b)$$
Using $U_0(0)=1$ and $U_1(0)=10^-4,$ the $s$ dependence of the solutions is 
depicted in Fig.~4. We notice that the energy 
from the pump will get monotonically transferred to the scattered mode.
\vskip 0.3 truecm
\noindent{\large \bf B.\ \  Production of the anti-Stokes component by intersecting masers}
\par\noindent
 It is clear from the previous section that the downshifted component can be easily created in 
astrophysical conditions. However, the upshifted component can only be produced in the forward
Raman scattering, which requires brightness temperature of the order of $10^{25}$~K, and it is
impossible to reach in astrophysical masers.

We propose a special mechanism$^9$ by invoking an inhomogeneity in the plasma to explain the 
upshifted
features seen in the spectrum of NGC~4258. The possibility of enhancing the forward scattering
by the backward scattering in the plasma, which has parabolic density profile 
$^{14,15}$ or by coupling Raman scattering with Brillouin scattering in the 
backward scattering $^{16}$ has been known from the experiments on laser-plasma 
interactions in the laboratory plasma.

Suppose two maser beams are intersecting in the plasma. If the maser beams are intersecting
in such a way that they produce the same longitudinal wave at the same direction, then the
energy transfer can take place between the two maser beams. In this process, the longitudinal
wave produced by the backward scattering of the incident radiation is exactly frequency 
matched with the upshifted scattered wave produced by the other incident radiation. The threshold
condition for the backward Raman scattering is
$$W_{\rm thr}= \frac{\alpha \alpha_3}{c_1 c_3}\mid\!\! E_o\!\!\mid^2 \cos^2\psi_1 > 1. \eqno (41)$$

The frequencies of the two intersecting maser beams need not be exactly same. The threshold 
condition implies that the intersecting angle of two masers has an allowance of about $\pm 10^o$ because $\vec k_1$ and $\vec k_3$ have a beam angle of this order of magnitude.

In this case, the transfer equations become very similar to  Eqs.~(24)-(26) as
$$\frac{\partial E_1}{\partial t}+c \frac{\partial E_1}{\partial x_1}=
   -c_1 E_1+\a E_o E_3^*\cos\psi_1,\eqno (42)$$
$$\frac{\partial E_2'}{\partial t}+c \frac{\partial E_2'}{\partial x_2'}=
   -c_2 E_2'-\a E_o' E_3^*\cos\psi_2'\eqno (43)$$
and 
$$\frac{\partial E_3}{\partial t}+v_g \frac{\partial E_3}{\partial x_3}=
   -c_3 E_3+\a_3 (E_o E_1^*\cos\psi_1+E_2' {E'}_o^*\cos\psi_2'),\eqno (44)$$
where the primed quantities are for the second maser beam, and $\vec k_2'-\vec k_o'=\vec k_3.$
In this process, the perfect phase matching is achieved for both the upshifted and downshifted
components, and we can repeat the analysis similar to the former treatment (Sect.~III). The 
growth rate is obtained as
$$\gamma=\frac{\alpha\alpha_3}{c_3}\left (\frac{\mid\!\! E_o\!\!\mid^2}{c_1}\cos^2\psi_1-
\frac{\mid\!\! E_o'\!\!\mid^2}{c_1}\cos^2\psi_2'\right )c_o.\eqno (45)$$
For $\psi_1=\psi_2'=0,$ the threshold condition is 
$$ (\mid\!\! E_o\!\!\mid^2-\mid\!\! E_o'\!\!\mid^2)\frac{\alpha\alpha_3}{c_3}c_o > 1.\eqno (46)$$
Therefore in order to excite the upshifted component the intensity of the first incident
maser must be greater than the intensity of the second maser. 
\vskip 0.3 truecm
\noindent{\large \bf IV.\ \  RESTRICTIONS OF THE THEORY}\par\noindent
        We have not included the incoherency effect due to the incident maser 
radiation. The bandwidth and maser beam solid angle determine the
coherence scale
length. If we take maser beam solid angle to be $10^{-4}$ and the bandwidth
of 1 km/sec, we can evaluate the coherency length approximately as $L_c=\l
(\triangle \O)^{-1}$ or $L_c=\l c/\triangle v_o\approx 10^4$--10$^5\l ,$
where $\l$ is the wavelength of the maser radiation. The Landau damping time
of the longitudinal plasma wave is one order of magnitude higher than the
coherence time scale $(\sim 10^{-5}$~s) of the incident radiation.  Hence the
monochromatic approximation used in the present analysis seems to be a good
approximation.  The effect of incoherence on the growth rate of stimulated
Raman scattering is discussed by $^{11}.$
        The second order Raman scattering of side band modes and the 
intensity level of the incident radiation are the main factors in limiting
(saturating) the amplitude of the side band modes.
\par
\vskip 0.3 truecm
\noindent{\large \bf V.\ \  CONCLUSION}\par\noindent
        The threshold condition for backward stimulated Raman
scattering is satisfied in the case of NGC~4258~$^9$. If the
compact HII regions are located near the center of maser cloud, maser
beam intersections can occur in it and  the upshifted component can
also be produced when the maser beams interact by coupling with a longitudinal 
plasma.  The Figs.~2 and 3 indicate the length scales for the
amplification of wave amplitudes in thick and thin plasma media, respectively. 
The inhomogeneity in the plasma and bandwidth in the incident radiation can
increase the amplification length by one order of magnitude.
\par
        The growth rate of stimulated Raman scattering is very large in the
case of backward scattering than that in forward scattering. This lead to the
complexity in explaining the spectrum and time variability of extreme
features.
\par
        The red and blue shifted features could be independent and produced
by two separate masers. A crucial test for the stimulated Raman maser model
is finding the triplet with an equal separation and not finding a
coherent time variation for the upshifted and downshifted components.
\vskip 1.0 cm
\noindent {\bf Acknowledgments}
\par
        This work was supported by Japan Society for the Promotion of Science 
(JSPS) Japan, and Department of Science and Technology (DST) India under
the India-Japan co-operative research programme.
\vfill\eject
\parindent=0pt\everypar={\hangindent=0.5 cm}
$^1$ V. Krishan, Astrophys. Lett., {\bf 23}, 133 (1983)

$^2$ V. Krishan, Current Science, {\bf 64}, 301 (1993)

$^3$ V. Krishan, P. J. Wiita, Mon. Not. R. Astron. Soc., {\bf 246}, 597 (1990)

$^4$ V. Krishan, P. J. Wiita, Astrophys. J, {\bf 423}, 172 (1994)

$^5$ R. T. Gangadhara, V. Krishan,  Mon. Not. R. Astron. Soc., {\bf 256}, 111 (1992)

$^6$ C. Fernandez, G. Reinnisch, Astron. \& Astrophys., {\bf 67}, 163 (1978)

$^7$ M. Miyoshi, J. Moran,  J. Herrnstein,  L. Greenhill, N. Nakai,
P. Diamond, M. Inoue, Nature,  {\bf 373}, 127 (1995)

$^8$ C. J. McKinstrie, A. Simon, E. A. Williams, Phys. Fluids, {\bf 27}, 2738 (1984)

$^9$ S. Deguchi, Astrophys. J, {\bf 420}, 551 (1994)

$^{10}$ R. T. Gangadhara, V. Krishan, Astrophys. J, {\bf 415}, 505 (1993)

$^{11}$ K. L. Kruer, The laser plasma interactions, (Addison-Wesley 
        Pub. Comp., Redwood city,  California, 1988) p. 90

$^{12}$ R. T. Gangadhara, V. Krishan,  Astrophys. J, {\bf 440}, 116 (1995)

$^{13}$ D. Anderson, Physica Scripta, {\bf 13}, 117 (1976)

$^{14}$ P. Koch, E. A. Williams, Phys. Fluids, {\bf 27}, 2346 (1984)

$^{15}$ R. E. Turner, K. Estarbrook, R. P. Drake, E. A. Williams, H. N. Kornblum,
        W. L. Kruer,  E. M. Campbell, Phys. Rev. Lett., {\bf 57}, 1725 (1986)

\noindent $^{16}$ C. Labaune, H. A. Baldis, S. D. Baton, D. Pesme, T. Jalinaud, Phys. Rev. Lett.,
          {\bf 69}, 285  (1992)
\vfill\eject
\begcap
\cap Fig.~1 The scattering geometry of the incident maser radiation in the 
case of thick plasma (a) and thin plasma (b) media.
\cap Fig.~2 The incident wave intensity $I_o/I_c$ and scattered wave intensity $I_1/I_c$
vs $ s$ in the case thick plasma medium at different scattering angles $\theta=120^o,$ 
$140^o,$ $160^o$ and $180^o.$ $I_c$ is a normalization constant 
and the parameter $s=(2 \nu_c \o_{pe}/c\o_o) x_1.$

\cap Fig.~3 The incident wave intensity $I_o/I_c$ and scattered wave intensity 
$I_1/I_c$ vs $s$ in the case thin plasma medium
at different scattering angles $\theta=120^o,$ $140^o,$ $160^o$ and $180^o.$
$I_c$ is a normalization constant and the parameter $s=(2 \nu_c \o_{pe}/c\o_o) x_1.$

\cap Fig.~4 Energy transfer between the coupled modes ($U_o$ 
corresponding to the pump wave)  and the parameter $s=(2 \nu_c \o_{pe}/c\o_o) x_1.$

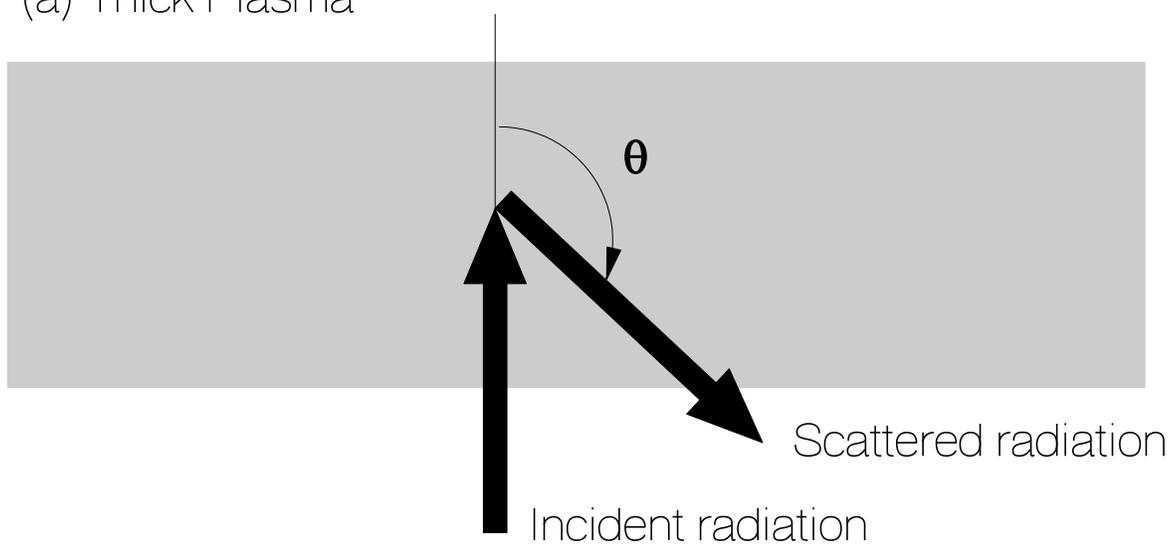
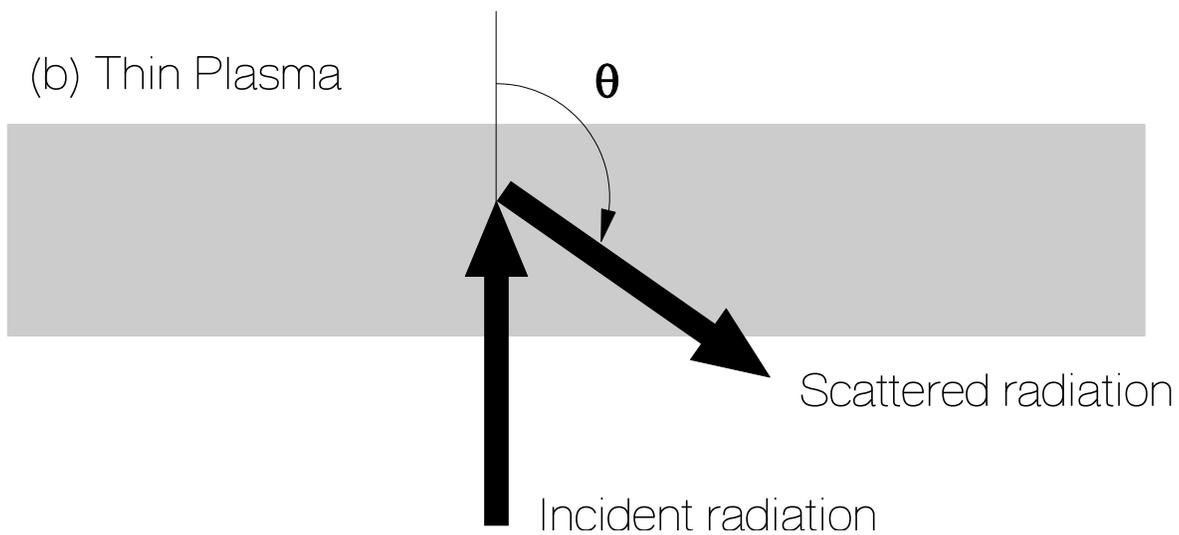
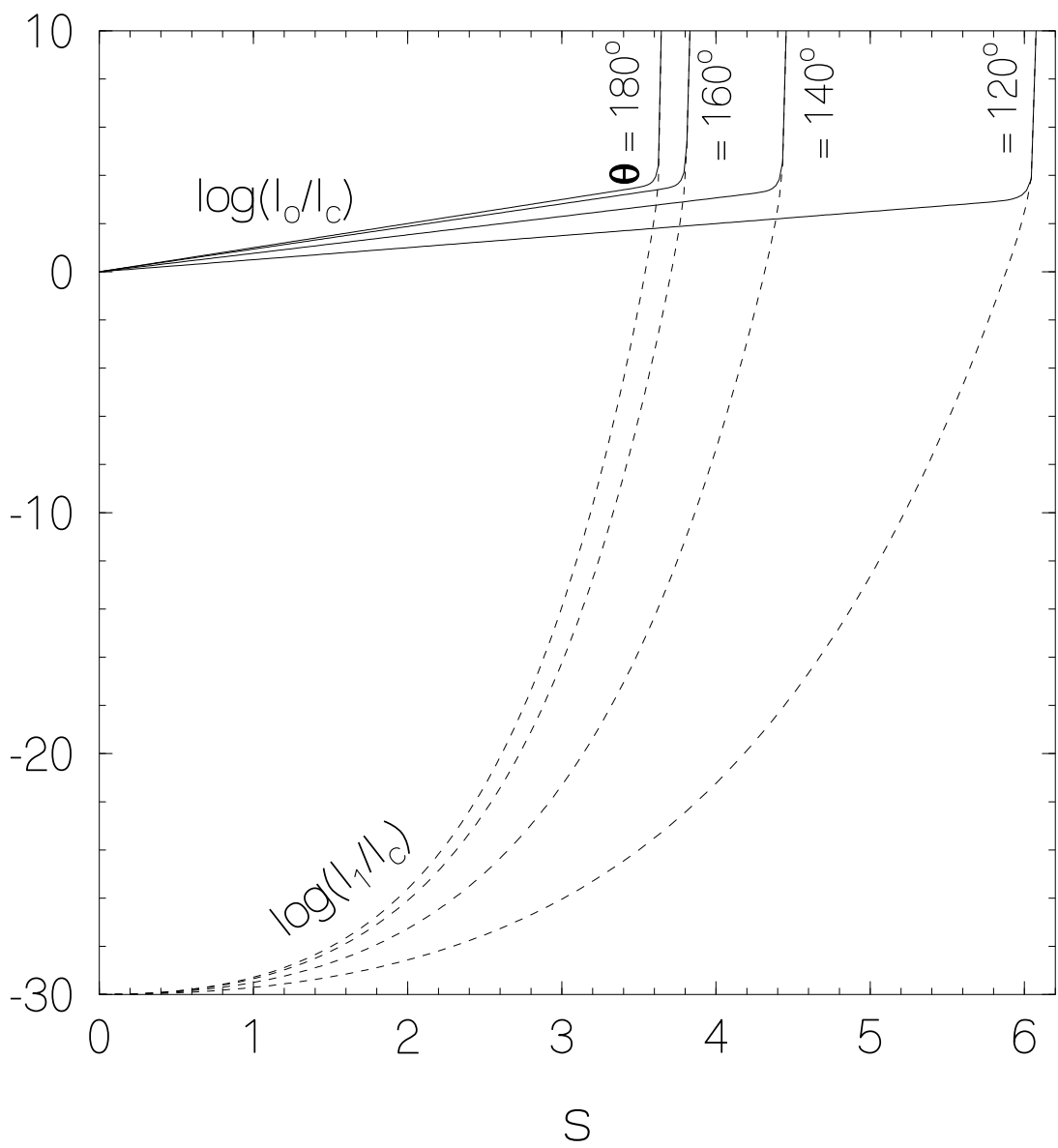
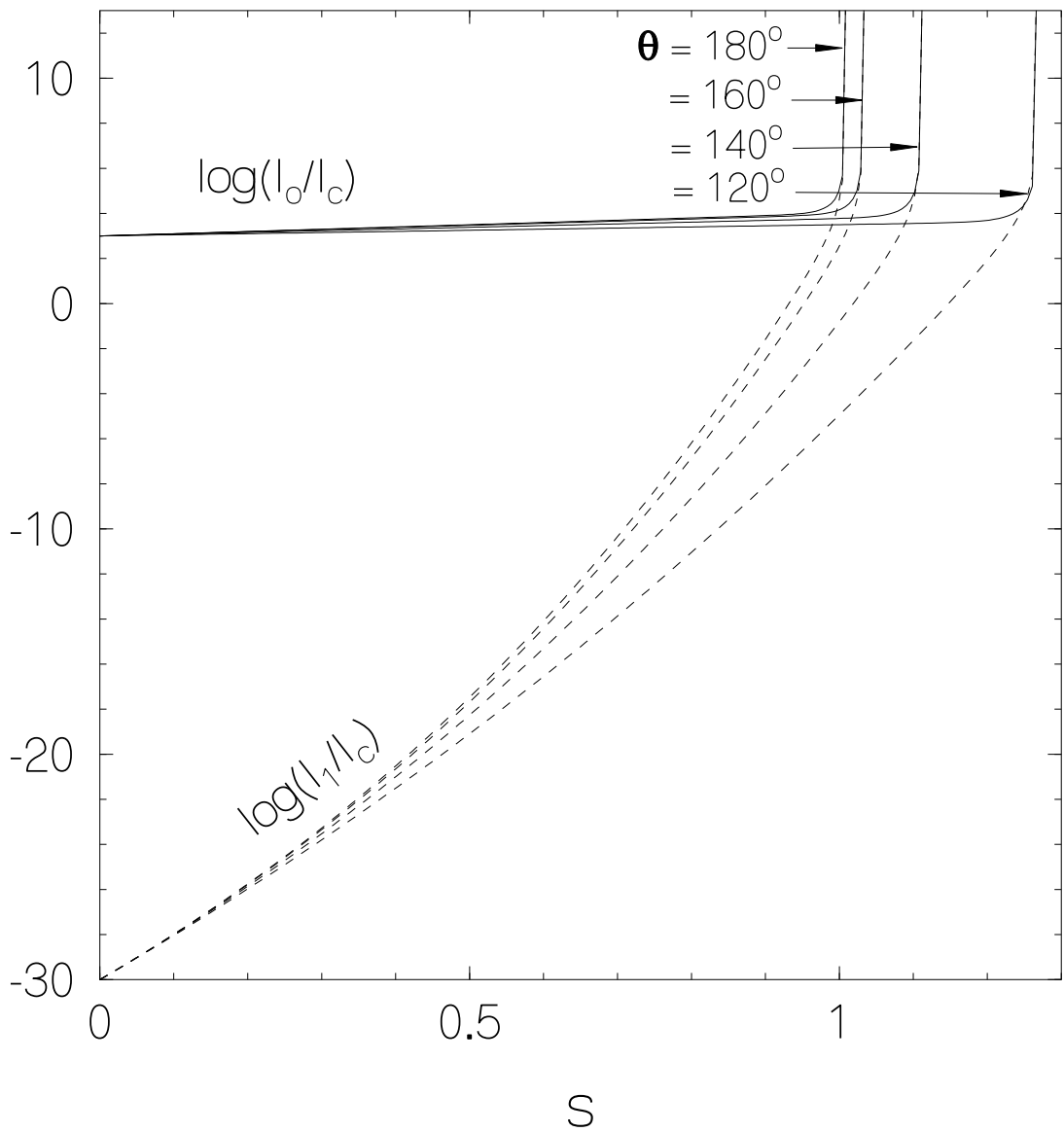
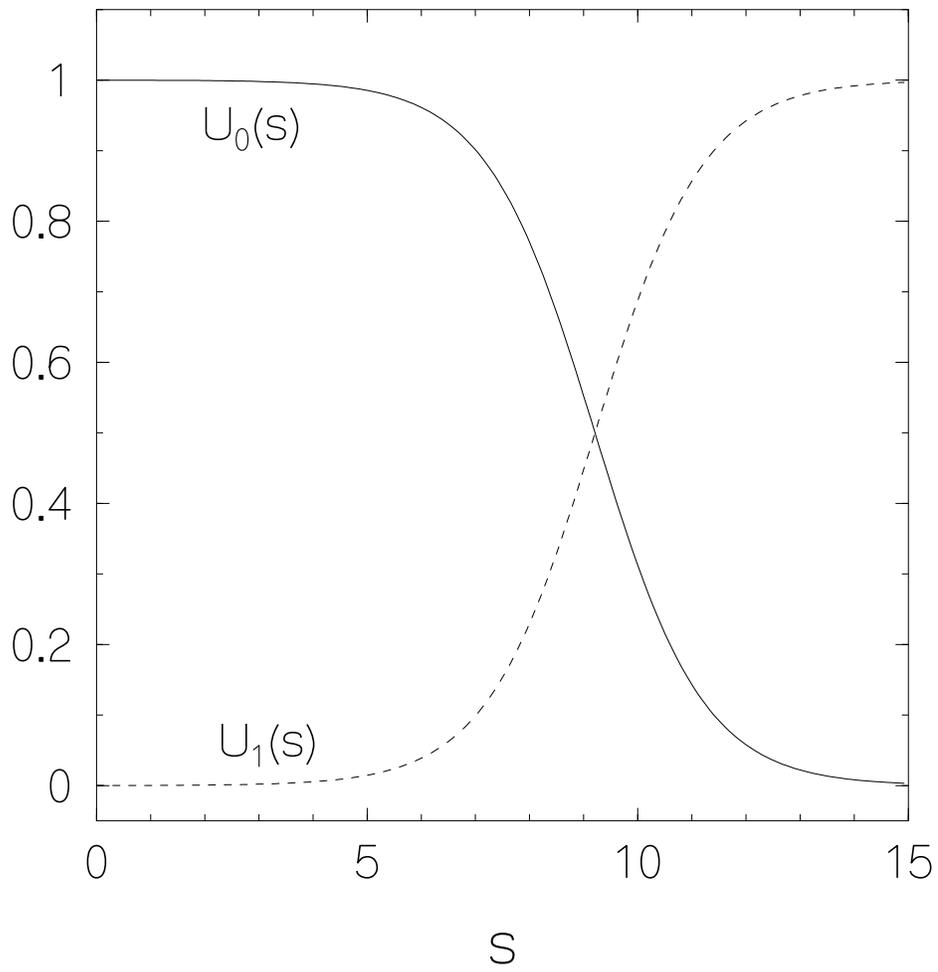
\end{document}